\documentclass[aps,prb,twocolumn,nobibnotes]{revtex4-1}

\usepackage[pdftex]{graphics}
\usepackage{amssymb,amsmath}
\usepackage{nicefrac}
\usepackage{physics}
\usepackage{color}
\usepackage{braket}
\usepackage{mathptmx}
\usepackage[T1]{fontenc}
\usepackage{bm}
\usepackage{fixmath}
\usepackage[normalem]{ulem}
\DeclareSymbolFont{boldoperators}{OT1}{cmr}{bx}{n}
\DeclareMathOperator{\acosh}{acosh}
\SetSymbolFont{boldoperators}{bold}{OT1}{cmr}{bx}{n}
\edef\bar{\unexpanded{\protect\mathaccentV{bar}}\number\symboldoperators16}
\mathchardef\mhyphen="2D
\begin{document}

\def\bra#1{\left<{#1}\right|}
\def\ket#1{\left|{#1}\right>}
\def\expval#1#2{\bra{#2} {#1} \ket{#2}}
\def\mapright#1{\smash{\mathop{\longrightarrow}\limits^{_{_{\phantom{X}}}{#1}_{_{\phantom{X}}}}}}

\title{An analysis of isomorphic RPMD in the golden rule limit}
\author{Joseph E. Lawrence}
\email[Email: ]{joseph.lawrence@chem.ox.ac.uk}
\affiliation{Department of Chemistry, University of Oxford, Physical and Theoretical\\ Chemistry Laboratory, South Parks Road, Oxford, OX1 3QZ, UK}
\author{David E. Manolopoulos}
\affiliation{Department of Chemistry, University of Oxford, Physical and Theoretical\\ Chemistry Laboratory, South Parks Road, Oxford, OX1 3QZ, UK}

\begin{abstract}
We analyse the golden rule limit of the recently proposed isomorphic ring polymer (iso-RP) method. This method aims to combine an exact expression for the quantum mechanical partition function of a system with multiple electronic states with a pre-existing mixed quantum-classical (MQC) dynamics approximation, such as fewest switches surface hopping. Since the choice of the MQC method adds a degree of flexibility, we simplify the analysis by assuming that the dynamics used correctly reproduces the exact golden rule rate for a non-adiabatic (e.g., electron transfer) reaction in the high temperature limit. Having made this assumption, we obtain an expression for the iso-RP rate in the golden rule limit that is valid at any temperature. We then compare this rate with the exact rate for a series of simple spin-boson models. We find that the iso-RP method does not correctly predict how nuclear quantum effects affect the reaction rate in the golden rule limit. Most notably, it does not capture the quantum asymmetry in a conventional (Marcus) plot of the logarithm of the reaction rate against the thermodynamic driving force, and it also significantly overestimates the correct quantum mechanical golden rule rate for activationless electron transfer reactions. These results are analysed and their implications discussed for the applicability of the iso-RP method to more general non-adiabatic reactions.
\end{abstract}

\maketitle
\section{Introduction}

The ability to simultaneously treat non-Born-Oppenheimer dynamics and nuclear quantum effects in atomistic simulations is important in a wide variety of contexts, ranging from proton coupled electron transfer reactions in proteins to energy transport in photovoltaics. Recently Tao, Shushkov and Miller have proposed a new method, called isomorphic ring polymer molecular dynamics (iso-RPMD or iso-RP),\cite{Tao18,Tao19} which aims to extend the well known Born-Oppenheimer ring polymer molecular dynamics (RPMD) method\cite{Craig04,Craig05a,Craig05b,Collepardo08,Boekelheide11,Habershon13} to systems with multiple electronic states. The iso-RPMD ansatz is to construct an ``isomorphic'' multi-state Hamiltonian for a ring polymer of classical nuclei that is consistent with the quantum mechanical partition function, and then to combine this with a pre-existing mixed quantum-classical (MQC) dynamics approximation such as Tully's fewest switches surface hopping.\cite{Tully90} 

One potential application for iso-PRMD is the study of thermal reactions which involve transitions between two different electronic states.\cite{Ghosh20} The most widely used theory for describing these reactions is Marcus theory,\cite{Marcus85} which assumes that the nuclear degrees of freedom in each state can be modelled with classical harmonic free energy surfaces and that the transitions between the states can be treated using perturbation theory (Fermi's golden rule). Marcus theory often works well at high temperatures, but it fails to capture nuclear quantum effects such as tunnelling and zero-point energy, which become important at lower temperatures. These nuclear quantum effects can be included in simple analytical theories,\cite{Kestner74} but in order to understand real chemical systems it is desirable to have a simulation method that includes nuclear quantum effects and is capable of treating fully atomistic models of multidimensional systems with anharmonicity. One might hope that iso-RPMD would provide a practical way to do this. 

In the adiabatic limit, in which excited electronic states do not make any contribution to the canonical quantum mechanical partition function, it is clear that iso-RPMD will work well, since it reduces in this limit to the standard RPMD approximation (in which the dynamics of the ring polymer takes place on the ground adiabatic Born-Oppenheimer potential energy surface). However, how well it can be expected to work in the opposite non-adiabatic limit, in which the rate of a transition between two diabatic electronic states is governed by the strength of the perturbative electronic coupling between these states as it is in Marcus theory, is less clear. This is what we shall investigate in the present paper.

\section{Isomorphic RPMD}

To set the scene for this investigation, let us begin by introducing the iso-RPMD formalism.\cite{Tao18,Tao19} Although this formalism is in principle applicable to systems with many electronic states, it will be sufficient for our purposes to consider just a two level system, for which the Hamiltonian can be written in the diabatic representation as
\begin{equation}
\hat{H}= \sum_{\nu=1}^{f} \frac{\hat{p}^2_\nu}{2m_\nu} \mathbf{1} + {\pmb{V}}(\hat{\pmb{q}})
\end{equation}
where $\mathbf{1}$ is the $2\times2$ identity matrix and ${\pmb{V}}(\hat{\pmb{q}})$ is the diabatic potential matrix
\begin{equation}
{\pmb{V}}(\hat{\pmb{q}})=\begin{pmatrix}
{V}_0(\hat{\pmb{q}}) & {\Delta}(\hat{\pmb{q}}) \\
{\Delta}(\hat{\pmb{q}}) & {V}_1(\hat{\pmb{q}})
\end{pmatrix}.
\end{equation}

The \lq\lq isomorphic\rq\rq\ RPMD for this two level system is based on writing the exact quantum partition function
\begin{equation}
Q = \Tr[e^{-\beta \hat{H}}] \label{exact_PF}
\end{equation} 
in the form
\begin{equation}
Q = \lim_{n\to\infty} \bigg(\frac{n}{2\pi\hbar}\bigg)^{nf} \int \mathrm{d}^{nf}\mathbf{p}\int\mathrm{d}^{nf} \mathbf{q}\,\tr_{\mathrm{e}}\Big[e^{-\beta \pmb{H}_n^{\mathrm{iso}}(\mathbf{p},\mathbf{q})}\Big],
\end{equation}
where $\tr_{\rm e}[\dots]$ denotes a trace over the electronic degrees of freedom. The isomorphic ring polymer Hamiltonian $\pmb{H}_n^{\mathrm{iso}}(\mathbf{p},\mathbf{q})$ is
\begin{equation}
\pmb{H}_n^{\mathrm{iso}}(\mathbf{p},\mathbf{q}) = h_n(\mathbf{p},\mathbf{q})  \mathbf{1}+\pmb{V}^{\mathrm{iso}}_n(\mathbf{q}),
\end{equation}
where
\begin{equation}
h_n(\mathbf{p},\mathbf{q}) = \sum_{j=1}^n\sum_{\nu=1}^f\Bigg\{ \frac{{p}^2_{j,\nu}}{2m_{n,\nu}} + \frac{1}{2}m_{n,\nu}\omega_{n}^2\big(q_{j+1,\nu}-q_{j,\nu}\big)^2 \Bigg\}
\end{equation}
is the free ring polymer Hamiltonian, with $m_{n,\nu}=m_{\nu}/n$, $\omega_n = n/(\beta\hbar)$, and $q_{n+1,\nu}=q_{1,\nu}$. The isomorphic diabatic potential energy matrix is
\begin{equation}
\pmb{V}^{\mathrm{iso}}_n(\mathbf{q})=\begin{pmatrix}
{V}^{\mathrm{iso}}_{0,n}(\mathbf{q}) & {\Delta}_n^{\mathrm{iso}}(\mathbf{q}) \\
{\Delta}_n^{\mathrm{iso}}(\mathbf{q}) & {V}^{\mathrm{iso}}_{1,n}(\mathbf{q})
\end{pmatrix},
\end{equation}
where the diagonal elements are chosen to be\cite{Tao18}
\begin{equation}
{V}^{\mathrm{iso}}_{i,n}(\mathbf{q}) = \frac{1}{n}\sum_{j=1}^n{V}_i(\pmb{q}_j).
\end{equation}
Here $\mathbf{q}=(\pmb{q}_1,\pmb{q}_2,\dots,\pmb{q}_n)$ is a vector whose elements are the position vectors for each ring polymer bead, and $\mathbf{p}=(\pmb{p}_1,\pmb{p}_2,\dots,\pmb{p}_n)$ are the conjugate momenta. 

In order to ensure that the iso-RP partition function exactly reproduces the quantum partition function in the infinite $n$ limit, Tao \emph{et al.}\cite{Tao18} require that $\Delta_n^{\mathrm{iso}}(\mathbf{q})$ is chosen so that
\begin{equation}
\tr_{\mathrm{e}}\Big[e^{-\beta \pmb{H}_n^{\mathrm{iso}}(\mathbf{p},\mathbf{q})}\Big]=e^{-\beta h_n(\mathbf{p},\mathbf{q})} \mu(\mathbf{q})
\end{equation}
where
\begin{equation}
\mu(\mathbf{q}) = \tr_{\mathrm{e}}\bigg[\mathcal{T}\prod_{j=1}^n e^{-\beta_n \pmb{V}(\pmb{q}_j)}\bigg],
\end{equation}
with $\beta_n=\beta/n$. Here $\mathcal{T}$ orders the matrix product in order of increasing bead index $j$, in accordance with a standard path integral discretisation of Eq.~(3). This leads to the following expression for the isomorphic coupling\cite{Tao18}
\begin{equation}
\Delta_n^{\rm iso}({\bf q})^2 = \acosh^2\left[{\mu({\bf q})\over 2}e^{\,\beta V_{+,n}^{\rm iso}({\bf q})/2}\right]/\beta^2
-V_{-,n}^{\rm iso}({\bf q})^2/4,
\end{equation}
where
\begin{equation}
V_{\pm,n}^{\rm iso}({\bf q}) = V_{0,n}^{\rm iso}({\bf q})\pm V_{1,n}^{\rm iso}({\bf q}).
\end{equation}

Note that in general $\mu(\mathbf{q})$ can become negative,\cite{Lu17} which leads to a complex $\Delta_n^{\mathrm{iso}}(\mathbf{q})$ and a non-hermitian $\pmb{V}^{\mathrm{iso}}_n(\mathbf{q})$. However, when ${\Delta}(\pmb{q})$ has a constant sign, one can show that  $\mu(\mathbf{q})>0$,\cite{Lu17} so $\Delta_n^{\mathrm{iso}}(\mathbf{q})$ is real and $\pmb{V}^{\mathrm{iso}}_n(\mathbf{q})$ is real and symmetric for all values of $\mathbf{q}$. We shall therefore assume  that $\Delta(\pmb{q})$ has a constant sign from this point on.

All of the above is simply a rearrangement of a formally exact expression for the quantum mechanical partition function. The isomorphic RPMD ansatz is to take the classical multi-state Hamiltonian $\pmb{H}_n^{\mathrm{iso}}(\mathbf{p},\mathbf{q})$ and use it with a mixed quantum classical (MQC) dynamics method, such as Tully's fewest switches surface hopping.\cite{Tully90} This is an appealingly simple idea, but how well does it work? In the following we will consider the more specific question of whether or not iso-RPMD can give accurate reaction rates in the non-adiabatic (golden rule) limit, and in particular whether it can capture nuclear quantum effects such as tunnelling and zero point energy in this limit.

\section{Golden Rule Limit}

The exact quantum mechanical rate constant for transfer from state $\ket{0}$ to state $\ket{1}$ in the golden rule limit can be written as\cite{Wolynes87}
\begin{equation}
k=\frac{1}{Q_r\hbar^2}\int_{-\infty}^{\infty} c(t) \mathrm{d}t \label{exact_rate},
\end{equation} 
where $Q_r$ is the reactant partition function and
\begin{equation}
c(t) = \tr_n\Big[e^{-\beta \hat{H}_0}e^{-i\hat{H}_0t/\hbar}{\Delta}(\hat{\pmb{q}})e^{+i\hat{H}_1t/\hbar}{\Delta}(\hat{\pmb{q}})\Big].
\end{equation}
Here
\begin{equation}
\hat{H}_i = \sum_{\nu=1}^{f} \frac{\hat{p}^2_\nu}{2m_\nu} + {V}_i(\hat{\pmb{q}})
\end{equation}
is the nuclear diabatic Hamiltonian on state $i$, and $\tr_n[\dots]$ denotes a trace over the nuclear coordinates. 

The classical limit of Eq.~(\ref{exact_rate}), which becomes exact at high temperatures where the ring polymer can be replaced by a single bead, is simply\cite{Sumi01}
\begin{equation}
k_{\mathrm{cl}}=\Big\langle\frac{2\pi}{\hbar}\Delta(\pmb{q})^2\delta\big(V_-(\pmb{q})\big)\Big\rangle_0 , \label{LZ_Rate}
\end{equation}
where $V_-(\pmb{q})=V_0(\pmb{q})-V_1(\pmb{q})$ is the diabatic energy gap and the thermal average is defined as
\begin{equation}
\langle A(\pmb{q}) \rangle_0 =\frac{\int \mathrm{d}^{f} \pmb{p} \int \mathrm{d}^{f} \pmb{q}\, e^{-\beta H_0(\pmb{p},\pmb{q})}A(\pmb{q})}{\int \mathrm{d}^{f} \pmb{p} \int \mathrm{d}^{f} \pmb{q}\, e^{-\beta H_0(\pmb{p},\pmb{q})}}.
\end{equation}

Unfortunately, it is well known that many MQC dynamics methods do not give the correct rate in the high-temperature, non-adiabatic limit (i.e.,\ are not consistent with Eq.~(\ref{LZ_Rate})). In particular, the original formulation of Tully's fewest switches surface hopping\cite{Tully90} suffers from ``over coherence''. There has been a concerted effort to fix this problem with decoherence corrections,\cite{Bittner95,Schwartz96,Prezhdo97,Fang99,Zhu04,Granucci07,Shenvi11,Landry11,Falk14} but that is not the focus of the present study. Here we shall simply sidestep the issue by assuming that the MQC method used with iso-RPMD {\em is} consistent with Eq.~(\ref{LZ_Rate}) in the one bead limit.

Having made this assumption, we need not consider the specific details of the MQC dynamics, and can simply make use of the isomorphism to write down the iso-RP expression for the golden rule rate
\begin{equation}
k_{\mathrm{iso}}=\lim_{n\to\infty}\Big\langle\frac{2\pi}{\hbar}\Delta^{\mathrm{iso}}_n(\mathbf{q})^2\delta\big(
V_{-,n}^{\mathrm{iso}}(\mathbf{q})\big)\Big\rangle^{\mathrm{iso}}_{0,n}, \label{iso_Rate}
\end{equation}
where the ring polymer thermal average is defined as
\begin{equation}
\langle A(\mathbf{q})\rangle_{0,n}^{\mathrm{iso}} =\frac{\int \mathrm{d}^{nf} \mathbf{p} \int \mathrm{d}^{nf} \mathbf{q}\,  e^{-\beta H^{\mathrm{iso}}_{0,n}(\mathbf{p},\mathbf{q})}A(\mathbf{q})}{\int \mathrm{d}^{nf} \mathbf{p} \int \mathrm{d}^{nf} \mathbf{q}\,  e^{-\beta H^{\mathrm{iso}}_{0,n}(\mathbf{p},\mathbf{q})}} , 
\end{equation}
with $H^{\mathrm{iso}}_{0,n}(\mathbf{p},\mathbf{q})=h_n(\mathbf{p},\mathbf{q})+V^{\mathrm{iso}}_{0,n}(\mathbf{q})$. 

We can further simplify Eq.~(\ref{iso_Rate}) by taking the golden rule limit of the iso-RP coupling $\Delta^{\mathrm{iso}}_n(\mathbf{q})^2$. To keep the notation as simple as possible, we shall do this only in the special case where $\hat{\Delta}(\pmb{q})=\Delta$ is a constant (the Condon approximation\cite{Nitzan06}). The more general case of a coordinate-dependent $\Delta$ can also be treated but it does not add anything useful to our discussion. The golden rule limit of $\Delta_n^{\rm iso}({\bf q})^2$ is worked out for constant $\Delta$ in Appendix~A, where it is shown to lead to a golden rule iso-RP rate in the infinite $n$ limit that can be written in path integral notation as
\begin{equation}
k_{\mathrm{iso}} =\frac{2\pi}{\beta\hbar Q_r}\int_0^\beta d\lambda \Big\langle \Delta^2 \delta\big(V_-^{\mathrm{iso}}[\pmb{q}(\tau)]\big)\Big\rangle_{\lambda} e^{-\beta F(\lambda)}. \label{K_ISO}
\end{equation}
Here the $\lambda$-dependent path-integral expectation value is defined as
\begin{equation}
\langle A[\pmb{q}(\tau)]\rangle_{\lambda} =\frac{\displaystyle{\oint}\mathcal{D}\pmb{q}(\tau)\, e^{-S^{(\lambda)}[\pmb{q}(\tau)]/\hbar}A[\pmb{q}(\tau)]}{\displaystyle{\oint}\mathcal{D}\pmb{q}(\tau)\, e^{-S^{(\lambda)}[\pmb{q}(\tau)]/\hbar}} , 
\end{equation}
with the action
\begin{equation}
S^{(\lambda)}[\pmb{q}(\tau)] = S^{(\lambda)}_0[\pmb{q}(\tau)] + S^{(\lambda)}_1[\pmb{q}(\tau)]
\end{equation}
\begin{equation}
S^{(\lambda)}_0[\pmb{q}(\tau)] = \int_{\lambda\hbar}^{\beta\hbar} \sum_{\nu=1}^f \frac{1}{2}m_\nu \dot{q}_\nu^2(\tau)+ V_0(\pmb{q}(\tau))\,\mathrm{d}\tau
\end{equation}
\begin{equation}
S^{(\lambda)}_1[\pmb{q}(\tau)] = \int_0^{\lambda\hbar} \sum_{\nu=1}^f \frac{1}{2}m_\nu \dot{q}_\nu^2(\tau)+ V_1(\pmb{q}(\tau))\,\mathrm{d}\tau,
\end{equation}
and the $\lambda$-dependent Boltzmann factor is
\begin{equation}
{e^{-\beta F(\lambda)}\over Q_r} = \frac{\displaystyle{\oint}\mathcal{D}\pmb{q}(\tau)\,e^{-S^{(\lambda)}[\pmb{q}(\tau)]/\hbar}}{\displaystyle{\oint}\mathcal{D}\pmb{q}(\tau)\,e^{-S^{(0)}[\pmb{q}(\tau)]/\hbar}}.
\end{equation}
The isomorphic diabatic energy gap in Eq.~(19) can be written in the same notation as
\begin{equation}
V_-^{\mathrm{iso}}[\pmb{q}(\tau)]=V^{\mathrm{iso}}_0[\pmb{q}(\tau)]-V^{\mathrm{iso}}_1[\pmb{q}(\tau)],
\end{equation}
where
\begin{equation}
V^{\mathrm{iso}}_i[\pmb{q}(\tau)]=\frac{1}{\beta\hbar}\int_0^{\beta\hbar}V_i(\pmb{q}(\tau))\,\mathrm{d}\tau
\end{equation}
and $\pmb{q}(\tau+\beta\hbar)=\pmb{q}(\tau)$ due to the cyclic boundary condition.

\section{Example Calculation}

In order to assess the accuracy of Eq.~(20), and facilitate our analysis, we shall consider the prototypical model of condensed phase electron transfer, the spin-boson model
\begin{equation}
V_0(\pmb{q}) = \sum_{\nu=1}^f \frac{1}{2}\omega_\nu^2 \Big(q_\nu + \frac{c_\nu}{\omega_\nu^2}\Big)^2 
\end{equation}
\begin{equation}
V_1(\pmb{q}) = \sum_{\nu=1}^f \frac{1}{2}\omega_\nu^2  \Big(q_\nu - \frac{c_\nu}{\omega_\nu^2}\Big)^2 -\epsilon,
\end{equation}
again with constant $\Delta$. The effect of the nuclear motion on the electronic dynamics of this model (and hence the rate of transfer from state 0 to state 1) is fully characterised by the spectral density
\begin{equation}
J(\omega) = \frac{\pi}{2} \sum_{\nu=1}^f \frac{c_\nu^2}{\omega_\nu} \delta(\omega-\omega_\nu),
\end{equation}
in which we have chosen to work in mass-scaled coordinates such that $m_\nu=1$.

Due to the simplicity of this problem it is possible to obtain closed form expressions for each of the terms appearing in Eq.~(20). The $\lambda$-dependent Boltzmann factor can be written as\cite{Weiss08}
\begin{equation}
\frac{e^{-\beta F(\lambda)}}{Q_r} = e^{-\beta \tilde{F}(\lambda)+\lambda\epsilon}
\end{equation}
with
\begin{equation}
\tilde{F}(\lambda)\! =\!\! \frac{4}{\pi}\!\int_0^\infty\!\!\frac{J(\omega)}{\beta\hbar\omega^2}\! \bigg[\frac{1-\cosh(\omega \lambda \hbar)}{\tanh(\omega\beta\hbar/2)}\! +\! \sinh(\omega \lambda\hbar)\bigg] \mathrm{d}\omega.
\end{equation}
Furthermore, using the fact that $V_-^{\mathrm{iso}}[\pmb{q}(\tau)]$ only depends on the average (centroid) of the imaginary time path, $\bar{\pmb{q}}=\frac{1}{\beta\hbar}\int_0^{\beta\hbar}\pmb{q}(\tau)\,\mathrm{d}\tau$, it is straightforward to show that
\begin{equation}
\Big\langle \delta\big(V_-^{\mathrm{iso}}[\pmb{q}(\tau)]\big)\Big\rangle_{\lambda} = \frac{1}{2}\sqrt{\frac{\beta}{\pi\Lambda}} e^{-\beta\frac{(\Lambda-\epsilon)^2}{4\Lambda}+\Lambda\lambda\big(1-\frac{\lambda}{\beta}\big)-\lambda\epsilon}, \label{prefactor}
\end{equation}
where 
\begin{equation}
\Lambda=\frac{4}{\pi}\int_0^\infty \frac{J(\omega)}{\omega} \mathrm{d}\omega
\end{equation}
is the Marcus theory reorganisation energy. 

\begin{figure}[t]
 \resizebox{1.0\columnwidth}{!} {\includegraphics{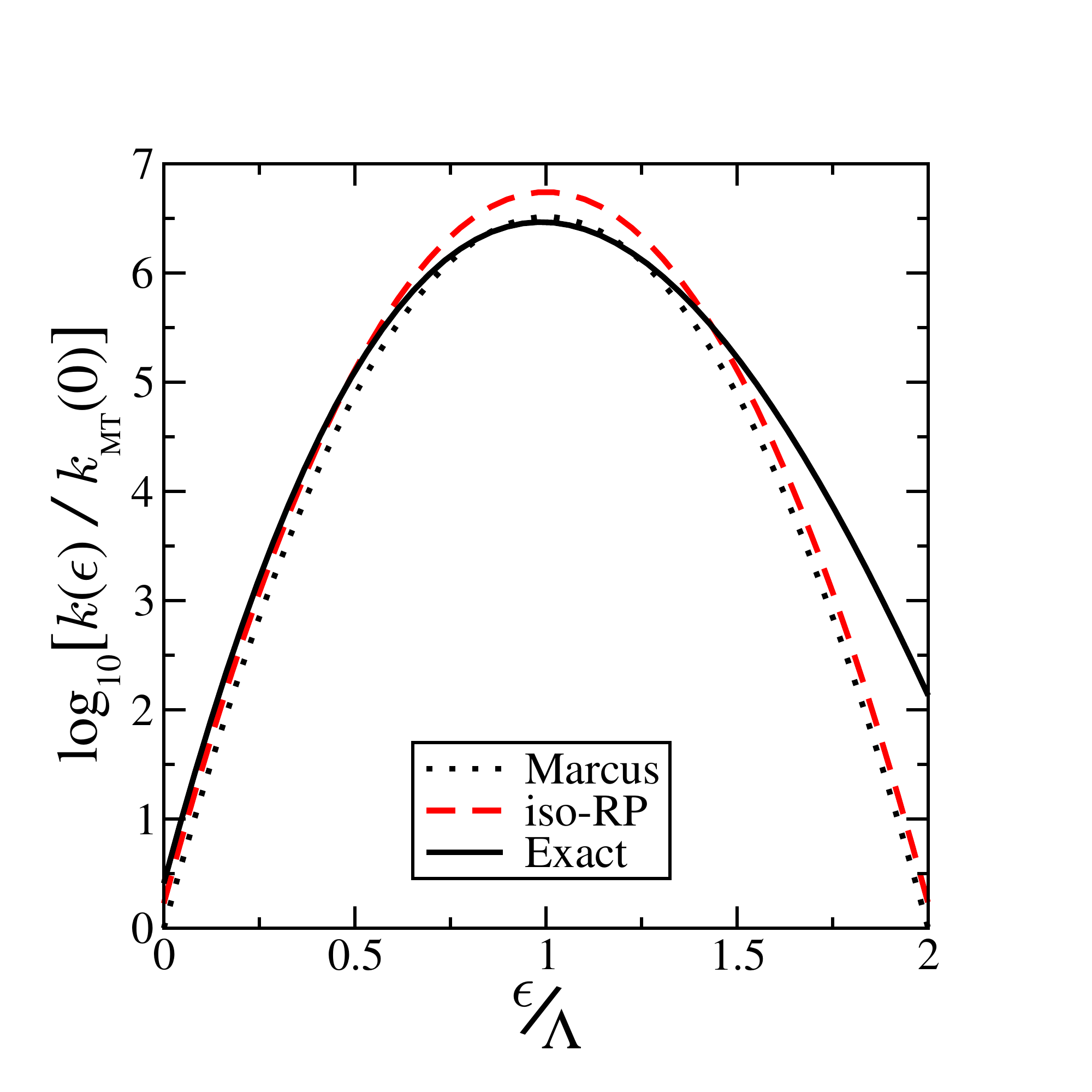}}
 \centering
 \caption{Reaction rate constants as a function of the driving force for model 1, a spin-boson model with $\beta\Omega=2$, $\gamma=\Omega$, and $\beta\Lambda=60$. Although it may seem as though the curvature of the iso-RP curve is larger (more negative) than that of the Marcus curve, this is an optical illusion: the logarithm of the iso-RP rate is in fact just vertically shifted from the Marcus theory parabola.}
 \label{SB_Rates}
 \end{figure}
 \begin{figure}[t]
 \resizebox{1.0\columnwidth}{!} {\includegraphics{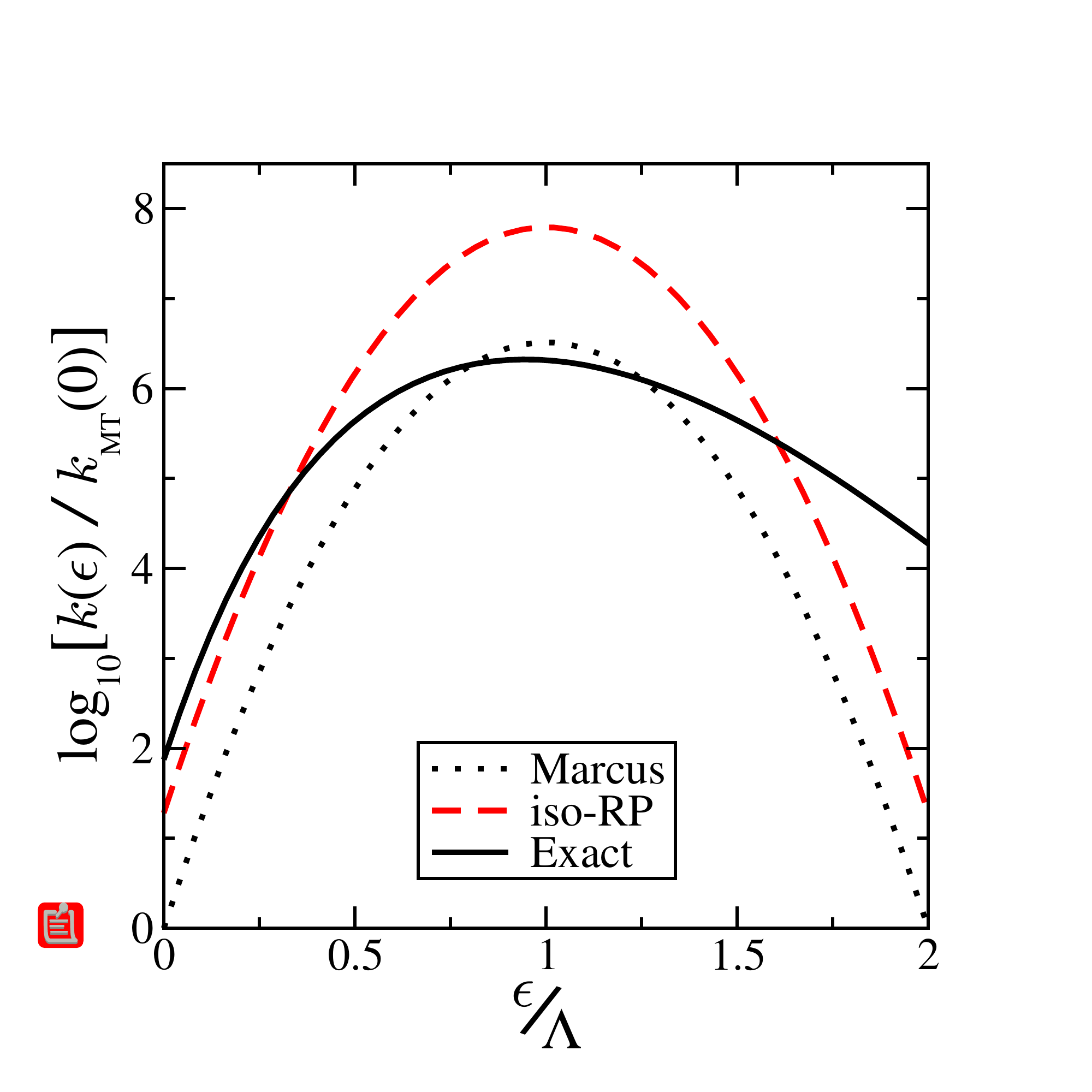}}
 \centering
 \caption{As in Fig.~1 but for model 2, a more strongly quantum mechanical spin-boson model with $\beta\Omega=6$, $\gamma=\Omega$, and $\beta\Lambda=60$.}
 \label{SB_Rates2}
 \end{figure}

Note that in the high temperature limit $\tilde{F}(\lambda)\to\Lambda\frac{\lambda}{\beta}\big(1-\frac{\lambda}{\beta}\big)$, and hence the $\lambda$ dependence of Eq.~(31) exactly cancels the $\lambda$ dependence of Eq.~(33). In this limit the isomorphic rate correctly reduces to the rate given by Marcus theory: $k_{\mathrm{iso}}\to k_{\mathrm{MT}}$ as $\beta\to0$ where\cite{Marcus85}
\begin{equation}
k_{\mathrm{MT}} = \frac{\Delta^2}{\hbar} \sqrt{\frac{\beta \pi}{\Lambda}} e^{-\beta\frac{(\Lambda-\epsilon)^2}{4\Lambda}}.
\end{equation}
Unfortunately, however, it is already clear that there is an issue with the iso-RP rate at lower temperatures. Since $\tilde{F}(\lambda)$ is independent of $\epsilon$ we see that the iso-RP rate retains the Marcus theory dependence on $\epsilon$ at all temperatures. Hence the ratio of the iso-RP rate to the Marcus rate is independent of the driving force, $k_{\mathrm{iso}}/k_{\mathrm{MT}}=A$, and so on a plot of the logarithm of the rate as a function of the driving force the iso-RP rate is simply a vertically shifted version of the famous Marcus parabola. This clearly does not capture the behaviour of the correct quantum mechanical rate at low temperatures.

To demonstrate this, we shall consider two models with Brownian Oscillator spectral densities\cite{Garg85,Leggett84,Thoss01}
\begin{equation}
J(\omega) = \frac{\Lambda}{2}\frac{\gamma \Omega^2 \omega}{(\omega^2-\Omega^2)^2+\gamma^2\omega^2},
\end{equation}
and with parameters chosen to be representative of typical condensed phase electron transfer reactions. Model 1 is weakly quantum mechanical with $\beta\Omega=2$, $\gamma=\Omega$, and $\beta\Lambda=60$, corresponding to a reaction coordinate frequency of $\Omega\approx\ 400$ cm$^{-1}$ and a Marcus reorganisation energy of $\Lambda\approx1.5$ eV at 300 K. Model 2 is more strongly quantum mechanical with $\beta\Omega=6$, $\gamma=\Omega$, and $\beta\Lambda=60$, corresponding to a reaction coordinate frequency of $\Omega\approx1200$ {cm}$^{-1}$. These parameters are roughly comparable to those obtained in the classic experimental paper on the Marcus inverted regime by Miller {\em et al.},\cite{Miller84} who extracted 1500 {cm}$^{-1}$ as the characteristic inner sphere frequency, along with a total reorganisation energy of 1.2 eV, for a series of organic electron transfers in tetrahydrofuran. The approximately 40$\%$ to 60$\%$ split of the reorganisation energy into inner-sphere and outer-sphere contributions they obtained experimentally is also broadly compatible with the Brownian Oscillator spectral density for model 2, in which $60\%$ of the reorganisation energy comes from modes with frequencies less than $\omega\approx1000$  {cm}$^{-1}$ at 300 K.

 Figure~\ref{SB_Rates} compares the iso-RP rate for model 1 as a function of the driving force, $\epsilon$, with both the exact quantum mechanical rate and the classical rate (given by Marcus theory). The isomorphic rate is unable to capture the dependence of the exact rate on the driving force, retaining the parabolic form seen in Marcus theory and failing to reproduce the large increase in the rate in the inverted regime due to nuclear quantum effects. The largest error is in the inverted regime: when the bias is twice the reorganisation energy the iso-RP rate is too small by nearly 2 orders of magnitude. In contrast it is reasonably accurate in the case of zero bias ($\epsilon=0$), where $k_{\mathrm{iso}}/k_{\mathrm{MT}}=1.7$ compared to $k/k_{\mathrm{MT}}=2.6$ for the exact rate. Perhaps the most worrying aspect of the results in  Fig.\ \ref{SB_Rates}, however, is that the iso-RP rate seems to be significantly less accurate than the Marcus  rate in the activationless case ($\epsilon=\Lambda$), where it predicts a quantum enhancement whereas the exact rate is actually slightly smaller than the Marcus rate. 
 
The overestimation of the rate in the activationless case is even more striking in model 2, as shown in Fig.~\ref{SB_Rates2}. Here the iso-RP rate is over 30 times larger than the exact rate at $\epsilon=\Lambda$. Moreover the error is again largest deep in the inverted regime, with the iso-RP rate underestimating the exact rate by nearly 3 orders of magnitude at $\epsilon=2\Lambda$, and it is again smallest for symmetric electron transfer ($\epsilon=0$), where the iso-RP rate is a factor of 4 too small. 
 
 \section{Analysis}
 
 To understand the behaviour of the iso-RP rate it is helpful to consider another approximate method for calculating reaction rates in the golden rule limit, namely Wolynes theory.\cite{Wolynes87} This theory makes a steepest descent approximation to the time integral in Eq.~(\ref{exact_rate}) to give
\begin{equation}
k_{\mathrm{WT}} = \frac{\Delta^2}{Q_r\hbar}\sqrt{\frac{2\pi}{-\beta F''(\lambda_{sp})}}e^{-\beta F(\lambda_{sp})}
\end{equation}
where the saddle point condition is $F'(\lambda_{sp})=0$. Wolynes theory is well known to be very accurate for the spin-boson model. It correctly recovers the Marcus theory rate in the high-temperature limit ($k_{\mathrm{WT}}\to k_{\mathrm{MT}}$ as $\beta\to0$), and it is also remarkably accurate at lower temperatures, where the path-integral distribution corresponding to $e^{-\beta F(\lambda_{sp})}$ is dominated by imaginary time paths close to the golden-rule instanton.\cite{Richardson15} For example, for our strongly quantum mechanical model 2 at zero bias, $k_{\mathrm{WT}}/k=0.98$, and in the activationless case $k_{\mathrm{WT}}/k=1.00$. 

 It is clear that one of the main differences between $k_{\mathrm{WT}}$ and $k_{\mathrm{iso}}$ is that in Wolynes theory the Boltzmann factor $e^{-\beta F(\lambda)}$ is evaluated at $\lambda=\lambda_{sp}$, whereas the iso-RP rate [Eq.~(20)] involves an integral over all values of $\lambda$ in the interval $[0,\beta]$. This difference can be used to explain the behaviour of the iso-RP rate. We begin by noting that the ($\epsilon$ independent) ratio of the iso-RP rate to the Marcus theory rate is given by
 \begin{equation}
 \frac{k_{\mathrm{iso}}}{k_{\mathrm{MT}}}=\frac{1}{\beta}\int_0^\beta \mathrm{d}\lambda \frac{e^{-\beta \tilde{F}(\lambda)}}{e^{-\beta \tilde{F}_{\mathrm{MT}}(\lambda)}}, \label{rate_ratio}
 \end{equation} 
where $\tilde{F}_{\mathrm{MT}}(\lambda)=\Lambda\frac{\lambda}{\beta}\Big(1-\frac{\lambda}{\beta}\Big)$ is the high temperature limit of $\tilde{F}(\lambda)$. Then since $k_{\mathrm{WT}}\to k_{\mathrm{MT}}$ as $\beta\to0$ we can write
\begin{equation}
\frac{k_{\mathrm{WT}}(\epsilon)}{k_{\mathrm{MT}}(\epsilon)} =  \sqrt{\frac{F_{\mathrm{MT}}''({\lambda}_{\mathrm{MT}})}{F''(\lambda_{sp})}}\frac{e^{-\beta \tilde{F}(\lambda_{sp})}}{e^{-\beta \tilde{F}_{{\mathrm{MT}}}({\lambda}_{\mathrm{MT}})}}
\end{equation}
where 
\begin{equation}
{\lambda}_{\mathrm{MT}}(\epsilon)=\frac{\beta(\Lambda-\epsilon)}{2\Lambda}
\end{equation}
is the high temperature limit of $\lambda_{sp}$. Now assuming that both $\lambda_{sp}(\epsilon)\approx\lambda_{\mathrm{MT}}(\epsilon)$ and $F''(\lambda_{sp})\approx F_{\mathrm{MT}}''({\lambda}_{\mathrm{MT}})$ continue to hold at lower temperatures (at least approximately), we see that
\begin{equation}
\frac{e^{-\beta \tilde{F}\big(\frac{\beta(\Lambda-\epsilon)}{2\Lambda}\big)}}{e^{-\beta \tilde{F}_{{\mathrm{MT}}}\big(\frac{\beta(\Lambda-\epsilon)}{2\Lambda}\big)}} \approx \frac{k_{\mathrm{WT}}(\epsilon)}{k_{\mathrm{MT}}(\epsilon)}\approx \frac{k(\epsilon)}{k_{\mathrm{MT}}(\epsilon)}. \label{wolynes_rate_ratio}
\end{equation}
Finally, making the substitution $\lambda=\lambda_{\mathrm{MT}}(\epsilon)$ to convert the integration variable from $\lambda$ to $\epsilon$ in Eq.~(38), and using the approximation in Eq.~(41), we can rewrite the quantum enhancement in the rate predicted by iso-RPMD as 
 \begin{equation}
 \frac{k_{\mathrm{iso}}}{k_{\mathrm{MT}}} \approx \frac{1}{2\Lambda}\int_{-\Lambda}^\Lambda \mathrm{d}\epsilon \frac{k(\epsilon)}{k_{\mathrm{MT}}(\epsilon)}. \label{average_rate}
 \end{equation}  
For both models considered above (model 1 and model 2), this approximation is within $1\%$ of the actual ratio, indicating that the assumptions in Eq.~(41) are valid and that this does indeed provide a useful way to understand the behaviour of the iso-RP rate.
 
 From Eq.~(42) we see that the quantum enhancement predicted by iso-RPMD can be interpreted as the average of the true quantum enhancement over the entire Marcus normal regime, $\epsilon\in[-\Lambda,\Lambda]$. This can be used to understand why, for example, iso-RPMD underestimates the rate in the symmetric case and overestimates the rate in the activationless case. Symmetric reactions (with $\epsilon=0$) have a larger quantum enhancement than other driving forces in the normal regime, due to nuclear tunnelling. The iso-RP rate for a symmetric reaction is therefore contaminated by imaginary time paths for which the tunnelling is weaker, which leads to an underestimation of the quantum mechanical rate. Conversely in the activationless case, where the exact rate is typically very close to the Marcus theory rate, iso-RPMD considerably overestimates the rate as it is dominated by the large nuclear tunnelling seen around $\epsilon=0$. In both cases, iso-RPMD is dominated by unphysical imaginary time paths, which leads to an inaccurate rate. This is in contrast to the situation in the adiabatic limit, where RPMD rate theory accurately captures deep tunnelling due to its connection to the semiclassical instanton approximation.\cite{Richardson09} Modifying iso-RPMD so that it is dominated by the correct (non-adiabatic) instanton path in the golden rule limit may well therefore provide a useful route to improving its accuracy for non-adiabatic reactions. \color{black}
  
\section{Conclusion}

Our analysis has shown that the iso-RP ansatz is unable to quantitatively describe reaction rates in the golden rule limit, when nuclear quantum effects are important. We have shown that the ansatz leads to a uniform quantum enhancement over the Marcus theory rate, independent of the thermodynamic driving force. Hence it does not capture the quantum asymmetry that is seen in a plot of the logarithm of the rate constant against the thermodynamic driving force. (However, since it correctly describes the high temperature limit, iso-RP will capture the classical asymmetry seen in anharmonic systems.) Furthermore, we have shown that the quantum enhancement predicted by the iso-RP ansatz is approximately the average of the true quantum enhancement over the entire Marcus normal regime. This explains why iso-RP tends to underestimate the rate for symmetric reactions and to overestimate the rate for activationless reactions. Since the average quantum enhancement in the normal regime is typically dominated by reactions near the symmetric limit, iso-RPMD may prove qualitatively useful in describing these reactions. However, it is clear that applying the method to reactions with driving forces nearer the activationless limit, or those in the Marcus inverted regime, would be inadvisable.

By design we have only considered the golden rule limit of iso-RPMD and it is clear that the failings seen in this regime will not be universal to all coupling strengths. In particular, provided a sensible MQC method is used, iso-RPMD should still accurately describe systems which are close to being electronically adiabatic. Furthermore, even for systems in which there are significant non-adiabatic effects, provided these effects are not rate limiting in the process considered, iso-RPMD may provide accurate results. It does share the single most important feature of standard adiabatic RPMD, in that it obeys quantum detailed balance by construction (provided a MQC method with this property is used).\cite{Tao18,Tao19} We also expect that iso-RPMD will be accurate for systems in which the non-adiabatic transition is effectively classical and nuclear quantum effects are important for crossing a barrier away from this transition. In fact, the accuracy of iso-RPMD has already been demonstrated in such a situation by Tao \emph{et al.} in their calculation of state-resolved reaction rates for a one-dimensional two-state model of the $\mathrm{F}+\mathrm{H}_2$ reaction.\cite{Tao18,Tao19} Application of the method to similar but more complex systems will no doubt provide physical insight in the future.

For systems in which nuclear quantum effects are important and the rate limiting step involves passage through the through the diabatic crossing, such as many biological and inorganic electron transfer reactions, more work is clearly needed to make iso-RPMD an accurate method for calculating reaction rates. Provided such systems are in the golden rule limit, however, Wolynes theory already provides a quantitatively accurate way of calculating reaction rates in both the normal and inverted Marcus regimes.\cite{Wolynes87,Lawrence18} Moreover one can connect Wolynes theory in the golden rule limit to conventional RPMD rate theory in the Born-Oppenheimer limit using a simple interpolation formula, which enables the calculation of accurate reaction rates for arbitrary electronic coupling strengths.\cite{Lawrence19,Lawrence20}  In order to apply iso-RPMD to such systems one would need to modify the isomorphic Hamiltonian or the dynamics to make it more closely related to an accurate approximation to the quantum rate in Eq.~(\ref{exact_rate}) in the golden rule limit. This might be achieved by making a connection to Wolynes theory,\cite{Wolynes87} or to a newer path integral based approach such as the recently proposed golden rule quantum transition state theory of Thapa \emph{et al.},~\cite{Thapa19} which already has a very similar functional form to Eq.~(\ref{K_ISO}).  But whether or not such an improvement is possible remains to be seen.

\begin{acknowledgments}
We would like to thank T. F. Miller III, S. C. Althorpe, and J. O. Richardson for helpful comments on the first draft of this manuscript. J. E. Lawrence is supported by The Queen's College Cyril and Phillis Long Scholarship in conjunction with the Clarendon Fund of the University of Oxford and by the EPRSC Centre for Doctoral Training in Theory and Modelling in the Chemical Sciences, EPSRC grant no. EP/L015722/1.
 \end{acknowledgments}

\appendix
\section{Golden Rule Limit of $\Delta^{\mathrm{iso}}_n(\mathbf{q})^2$} 

Here we evaluate the golden rule limit of the iso-RP coupling, $\Delta_n^{\rm iso}({\bf q})$. In order to keep the argument as simple as possible, we will consider a position independent $\Delta(\pmb{q})=\Delta$. It is also simpler to work in path integral notation, where ring polymer configurations $\mathbf{q}$ (in the infinite $n$ limit) become cyclic paths $\pmb{q}(\tau)$ parameterised by imaginary time $\tau$. Consequently functions of a single ring polymer configuration, $A(\mathbf{q})$, become functionals of the imaginary time path, $A[\pmb{q}(\tau)]$. Making this notational change we can recast Eq.~(\ref{iso_Rate}) in the infinite $n$ limit as
\begin{equation}
k_{\mathrm{iso}}=\Big\langle\frac{2\pi}{\hbar}(\Delta^{\mathrm{iso}}[\pmb{q}(\tau)])^2\delta\big(V_-^{\mathrm{iso}}[\pmb{q}(\tau)]\big)\Big\rangle^{\mathrm{iso}}_{0}, \label{path_int_iso_Rate}
\end{equation}
where
\begin{equation}
V_-^{\mathrm{iso}}[\pmb{q}(\tau)]=V^{\mathrm{iso}}_0[\pmb{q}(\tau)]-V^{\mathrm{iso}}_1[\pmb{q}(\tau)]
\end{equation}
\begin{equation}
V^{\mathrm{iso}}_i[\pmb{q}(\tau)]=\frac{1}{\beta\hbar}\int_0^{\beta\hbar}V_i(\pmb{q}(\tau))\,\mathrm{d}\tau,
\end{equation}
and the expectation value is given by
\begin{equation}
\langle A[\pmb{q}(\tau)]\rangle_{0}^{\mathrm{iso}} =\frac{\displaystyle{\oint}\mathcal{D}\pmb{q}(\tau)\, e^{-S^{\mathrm{iso}}_0[\pmb{q}(\tau)]/\hbar}A[\pmb{q}(\tau)]}{\displaystyle{\oint}\mathcal{D}\pmb{q}(\tau)\, e^{-S^{\mathrm{iso}}_0[\pmb{q}(\tau)]/\hbar}} , \label{iso_expectation}
\end{equation}
with
\begin{equation}
S^{\mathrm{iso}}_0[\pmb{q}(\tau)] = \int_0^{\beta\hbar} \sum_{\nu=1}^f \frac{1}{2}m_\nu \dot{q}_\nu^2(\tau)+V_0(\pmb{q}(\tau))\,\mathrm{d}\tau.
\end{equation}
Note that the integrals over the positions of the ring polymer beads have become a ``path integral'' over all possible paths $\pmb{q}(\tau)$. The circle indicates that the paths are cyclic, $\pmb{q}(\tau+\beta\hbar)=\pmb{q}(\tau)$, and the measure $\mathcal{D}\pmb{q}(\tau)$ contains the integrals over the ring polymer momenta (which can be done analytically). 

In this notation, the isomorphic coupling in Eq.~(11) becomes
\begin{equation}
\begin{split}
\Delta^{\rm iso}\left[\pmb{q}(\tau)\right]^2 &= \acosh^2\left[{\mu[\pmb{q}(\tau)]\over 2}e^{\,\beta V_+^{\rm iso}[\pmb{q}(\tau)]/2}\right]/\beta^2\\ &-V_-^{\rm iso}[\pmb{q}(\tau)]^2/4,
\end{split}
\end{equation}
where
\begin{equation}
V_{+}^{\rm iso}[\pmb{q}(\tau)] = V_0^{\rm iso}[\pmb{q}(\tau)]+V_1^{\rm iso}[\pmb{q}(\tau)]
\end{equation}
and
\begin{equation}
\mu[\pmb{q}(\tau)]=\tr_\mathrm{e}\bigg[\mathcal{T}\exp\bigg(-\frac{1}{\hbar}\int_0^{\beta\hbar}\pmb{V}(\pmb{q}(\tau))\,\mathrm{d}\tau\bigg)\bigg],
\end{equation}
in which the $\mathcal{T}$ is now an imaginary time ordering operator and $\pmb{V}(\pmb{q}(\tau))$ is just the diabatic potential matrix in Eq.~(2) evaluated at $\pmb{q}(\tau)$. 

To find the golden rule limit of $\Delta^{\mathrm{iso}}[\pmb{q}(\tau)]^2$, we first note that the only term in Eq.~(A6) which depends on the physical coupling, $\Delta$, is $\mu[\pmb{q}(\tau)]$. Hence we begin by expanding  $\mu[\pmb{q}(\tau)]$ to second order in $\Delta$ as
\begin{equation}
\mu[\pmb{q}(\tau)]=\mu_0[\pmb{q}(\tau)]+{\Delta^2}\mu_2[\pmb{q}(\tau)]+\mathcal{O}(\Delta^4). \label{mu_expansion1}
\end{equation}
In order to evaluate each term in this expansion one can expand the trace as
\begin{equation}
\mu[\pmb{q}(\tau)] = \lim_{n\to\infty}\,\prod_{j=1}^n\bigg( \sum_{\sigma_j=0}^{1} \left<\sigma_j\right| e^{-\beta_n \pmb{V}(\pmb{q}(\tau_j))}\left|\sigma_{j+1}\right>\bigg),
\end{equation}
where $\tau_j=j\beta_n\hbar$ and $\sigma_{n+1}=\sigma_1$. Then noting that
\begin{equation}
 \left<0\right| e^{-\beta_n \pmb{V}(\pmb{q})}\left|0\right> = e^{-\beta_n V_0(\pmb{q})}+\mathcal{O}(\beta_n^2)
 \end{equation}
 \begin{equation}
 \left<1\right| e^{-\beta_n \pmb{V}(\pmb{q})}\left|1\right> = e^{-\beta_n V_1(\pmb{q})}+\mathcal{O}(\beta_n^2)
 \end{equation}
 and
\begin{equation}
 \left<0\right| e^{-\beta_n \pmb{V}(\pmb{q})}\left|1\right> = -\beta_n\Delta e^{-\beta_n [V_0(\pmb{q})+V_1(\pmb{q})]/2}+\mathcal{O}(\beta_n^2),
 \end{equation}
we see that each off-diagonal matrix element in the product in Eq.~(A10) contributes one power of $\Delta$. Equivalently, noting that the set $\{\sigma_j\}$ in the infinite $n$ limit corresponds to a stochastic path $\sigma(\tau)$, we see that every time $\sigma(\tau)$ changes from $0$ to $1$ (known as a kink) we pick up one power of $\Delta$. Since the path $\sigma(\tau)$ has cyclic symmetry it must have an even number of kinks and hence $\mu[\pmb{q}(\tau)]$ is an even function of $\Delta$. 

It is immediately obvious that the only paths which contribute to the zeroth order term are those which do not contain any kinks
\begin{equation}
\mu_0[\pmb{q}(\tau)] = e^{-\beta V^{\mathrm{iso}}_0[\pmb{q}(\tau)]}+e^{-\beta V^{\mathrm{iso}}_1[\pmb{q}(\tau)]},
\end{equation}
corresponding to $\sigma(\tau)=0$ and $\sigma(\tau)=1$ for all $\tau$ in the trace over diabatic states in Eq.~(A8). Similarly the only paths which contribute to $\mu_2[\pmb{q}(\tau)]$ are those which contain two kinks. Each of these paths can be characterised by two parameters, the total time which the path spends on diabat $1$, which we denote $\lambda\hbar$
\begin{equation}
\lambda\hbar = \int_0^{\beta\hbar} \sigma(\tau)\,  \mathrm{d}\tau,
\end{equation} 
and the time at which $\sigma(\tau)$ changes from $0$ to $1$, which we label $\tau'$. We can thus write the second order term as
\begin{equation} 
\mu_2[\pmb{q}(\tau)]\!=\!{1\over\hbar}\int_0^{\beta}\!\!\mathrm{d}\lambda\int_0^{\beta\hbar}\!\!\mathrm{d}\tau' e^{-(\beta-\lambda) V^{\lambda}_0[\pmb{q}(\tau+\tau')]-\lambda V^{\lambda}_1[\pmb{q}(\tau+\tau')]},
\end{equation}
where
\begin{align}
V^{\lambda}_0[\pmb{q}(\tau)]=&\frac{1}{(\beta-\lambda)\hbar}\int_{\lambda\hbar}^{\beta\hbar} V_0(\pmb{q}(\tau))\,\mathrm{d}\tau\\
V^{\lambda}_1[\pmb{q}(\tau)]=&\frac{1}{\lambda\hbar}\int_{0}^{\lambda\hbar} V_1(\pmb{q}(\tau))\,\mathrm{d}\tau.
\end{align}

Now by first recognising that
\begin{equation}
{\mu_0[\pmb{q}(\tau)]\over 2}e^{\,\beta V_+^{\rm iso}[\pmb{q}(\tau)]/2}=\cosh(\beta V^{\mathrm{iso}}_-[\pmb{q}(\tau)]/2),
\end{equation}
and then making use of 
\begin{equation}
\acosh^2[\cosh(a)+b\Delta^2]-a^2= \frac{2b\Delta^2 a}{\sinh(a)}+\mathcal{O}(\Delta^4),
\end{equation}
it is straightforward to show that we can write the coupling as
\begin{equation}
\Delta^{\mathrm{iso}}[\pmb{q}(\tau)]^2 = \Delta^2 \frac{\mu_2[\pmb{q}(\tau)]}{\mu^{\mathrm{iso}}_2[\pmb{q}(\tau)]}+\mathcal{O}(\Delta^4),\label{NAL_coupling}
\end{equation}
where
\begin{equation}
\mu^{\mathrm{iso}}_2[\pmb{q}(\tau)] = \frac{2\beta\sinh({\beta V^{\mathrm{iso}}_-[\pmb{q}(\tau)]/2})}{e^{\,\beta V^{\mathrm{iso}}_+[\pmb{q}(\tau)]/2}\,V^{\mathrm{iso}}_-[\pmb{q}(\tau)]},
\end{equation}
or equivalently
\begin{equation}
\mu^{\mathrm{iso}}_2[\pmb{q}(\tau)]\!=\!{1\over\hbar}\int_0^{\beta}\!\!\mathrm{d}\lambda\int_0^{\beta\hbar}\!\!\mathrm{d}\tau' e^{-(\beta-\lambda) V^{\mathrm{iso}}_0[\pmb{q}(\tau)]-\lambda V^{\mathrm{iso}}_1[\pmb{q}(\tau)]}.
\end{equation}
Next recognising that
\begin{equation}
\frac{\delta\big(V^{\mathrm{iso}}_{-}[\pmb{q}(\tau)]\big)}{\mu_2^{\mathrm{iso}}[\pmb{q}(\tau)]} = \frac{\delta\big(V^{\mathrm{iso}}_{-}[\pmb{q}(\tau)]\big)}{e^{-\beta V^{\mathrm{iso}}_0[\pmb{q}(\tau)]}\beta^2}, 
\end{equation}
and also that 
\begin{equation}
\frac{e^{-S^{\mathrm{iso}}_0[\pmb{q}(\tau)]/\hbar}\mu_2[\pmb{q}(\tau)]}{ e^{-\beta V^{\mathrm{iso}}_0[\pmb{q}(\tau)]}}=\!{1\over\hbar}\int_0^{\beta}\!\!\mathrm{d}\lambda\int_0^{\beta\hbar}\!\!\mathrm{d}\tau' e^{-S^{(\lambda)}[\pmb{q}(\tau+\tau')]/\hbar},
\end{equation}
where
\begin{equation}
S^{(\lambda)}[\pmb{q}(\tau)] = S^{(\lambda)}_0[\pmb{q}(\tau)] + S^{(\lambda)}_1[\pmb{q}(\tau)]
\end{equation}
\begin{equation}
S^{(\lambda)}_0[\pmb{q}(\tau)] =\int_{\lambda\hbar}^{\beta\hbar} \sum_{\nu=1}^f \frac{1}{2}m_\nu \dot{q}_\nu^2(\tau)+ V_0(\pmb{q}(\tau))\mathrm{d}\tau
\end{equation}
\begin{equation}
S^{(\lambda)}_1[\pmb{q}(\tau)] =\int_0^{\lambda\hbar} \sum_{\nu=1}^f \frac{1}{2}m_\nu \dot{q}_\nu^2(\tau)+ V_1(\pmb{q}(\tau))\mathrm{d}\tau,
\end{equation}
we see that we can rewrite Eq.~(\ref{path_int_iso_Rate}) in the golden rule limit as in Eq.~(25),
\begin{equation}
k_{\mathrm{iso}} =\frac{2\pi}{\beta\hbar Q_r}\int_0^\beta d\lambda \Big\langle \Delta^2 \delta\big(V_-^{\mathrm{iso}}[\pmb{q}(\tau)]\big)\Big\rangle_{\lambda} e^{-\beta F(\lambda)}, \label{K_ISO_appendix}
\end{equation}
with
\begin{equation}
\langle A[\pmb{q}(\tau)]\rangle_{\lambda} =\frac{\displaystyle{\oint}\mathcal{D}\pmb{q}(\tau) e^{-S^{(\lambda)}[\pmb{q}(\tau)]/\hbar}A[\pmb{q}(\tau)]}{\displaystyle{\oint}\mathcal{D}\pmb{q}(\tau) e^{-S^{(\lambda)}[\pmb{q}(\tau)]/\hbar}} 
\end{equation}
and
\begin{equation}
\begin{aligned}
e^{-\beta F(\lambda)}&=\tr_{n}[e^{-(\beta-\lambda)\hat{H}_0}e^{-\lambda \hat{H}_1}]\\&=\displaystyle{\oint}\mathcal{D}\pmb{q}(\tau)\, e^{-S^{(\lambda)}[\pmb{q}(\tau)]/\hbar}
\end{aligned}
\end{equation}
and 
\begin{equation}
\begin{aligned}
Q_r= e^{-\beta F(0)}&={\rm tr}_n[e^{-\beta \hat{H}_0}]\\
&=\displaystyle{\oint}\mathcal{D}\pmb{q}(\tau)\,e^{-S^{(0)}[\pmb{q}(\tau)]/\hbar}.
\end{aligned}
\end{equation}

\end{document}